\documentclass{ws-procs9x6}

\begin{document}

\title{NEUCAL: a prototype detector for electron/hadron discrimination through neutron measurement}

\author{L. BONECHI$^*$, O. ADRIANI, R. D'ALESSANDRO, P. SONA,\\G. SORICHETTI and P. SPILLANTINI}

\address{Physics Department of University of Florence and INFN Section of Florence,\\
50019 Sesto Fiorentino (Florence), Italy\\
$^*$E-mail: Lorenzo.Bonechi@fi.infn.it}

\author{S. BOTTAI, M. GRANDI, P. PAPINI, S. RICCIARINI, G. SGUAZZONI,\\E. VANNUCCINI and A. VICIANI}

\address{INFN Section of Florence,\\
50019 Sesto Fiorentino (Florence), Italy}

\author{G. CASTELLINI}

\address{IFAC-CNR\\
50019 Sesto Fiorentino (Florence), Italy\\}

\begin{abstract}
NEUCAL is a neutron detector which is currently under study to be used as a sub-detector complementing electromagnetic (e.m.) calorimeters for electron/hadron discrimination in cosmic rays at high energy. Its aim is to reveal the different yield of neutron production in e.m. and hadronic showers, not only by counting signals due to their absorption in some sensible detector after passive moderation, but also looking for signals produced during the moderation phase. The basic idea and a test of a prototype detector are discussed in this paper. A first preliminary comparison of experimental data with simulation is also shown.
\end{abstract}


\bodymatter

\section{Introduction}\label{sec:intro}
Electron/hadron discrimination is an important requirement for all HEP experiments detecting charged radiation and particularly for space experiments aiming to directly measure particle fluxes at the highest energies. The neutron detection technique has been recently introduced by the PAMELA\cite{pamela} satellite experiment to improve the e.m. calorimeter rejection capability, that is mainly based on the shower topology analysis. Interacting hadrons can be wrongly identified in the calorimeter as electrons, and this kind of events is a dangerous background especially for the cosmic-ray electron and positron flux measurements, since the hadron flux in primary cosmic rays is dominant. Significant differences for the neutron component leaking out from the bottom side of the calorimeter are anyway expected between e.m. and hadronic showers, as confirmed by simulations and experimental Pamela data; detecting these neutrons is the basic idea of NEUCAL. \newline
The aim of the NEUCAL detector is to improve neutron identification and counting, extending the e.m. calorimeter performance and reducing at the same time the request of heavy absorber material. Neutron detection is usually achieved by detecting the energy release due to the nuclear neutron capture, once the neutron has been moderated to reduce its kinetic energy around the thermal energy (E\,$\simeq$\,25\,meV). This requires using layers of light material like paraffin, polyethylene or other polymer, rich in hydrogen atoms. The new idea behind the development of the NEUCAL detector is to improve neutron measurements by detecting not only the signals due to the capture in $^3$He tubes, but also those rising during the moderation phase. Some ``active'' moderator material is therefore necessary, replacing the passive paraffin layers and capable of converting the energy releases into measurable signals. We have found that the standard fast plastic scintillators are particularly suitable for this kind of application, being rich in Hydrogen atoms (see for example polyvinyl-toluene, a polymer with chemical formula $CH_2CH(C_6H_4CH_3)_n$ commonly used for fast scintillators).

\section{Simulation of  hadronic and electromagnetic showers}\label{sec:simulation}
For a preliminary simulation of the NEUCAL detector we have considered as a possible application the CALET space experiment on-board the ISS, which is under discussion for final approval. The CALET experiment is based on a 32\,X$_0$ deep BGO calorimeter. NEUCAL was dimensioned as a twelve 1\,cm thick scintillator layers, interleaved by planes of 1\,cm diameter $^3$He proportional counter tubes. Samples of electron and proton events have been simulated at different energies and the shower development inside the calorimeter has been analyzed in details to study the spectra of neutrons exiting from its back and entering the NEUCAL detector. The first results\cite{bottai_elba} show that most of the neutrons exits from the calorimeter within one hundred nanoseconds, with energy between some tenths of MeV and one hundred MeV. The response of the NEUCAL detector to single neutrons has hence been studied in this energy range. Results show that neutrons around 1\,MeV are well moderated already in the first three centimeters of scintillator and can then be efficiently absorbed by the $^3$He tubes. In case of 10\,MeV neutrons only a fraction around $10\,\%$ is efficiently moderated while $30\,\%$ of them do not release detectable energy. The remaining $60\,\%$ releases energy amounts that can be detected in the scintillators.

\section{Production of the prototype detector}\label{sec:prototype}
\begin{figure}[top]
\begin{center}
\boxed{\includegraphics[height=2.6cm]{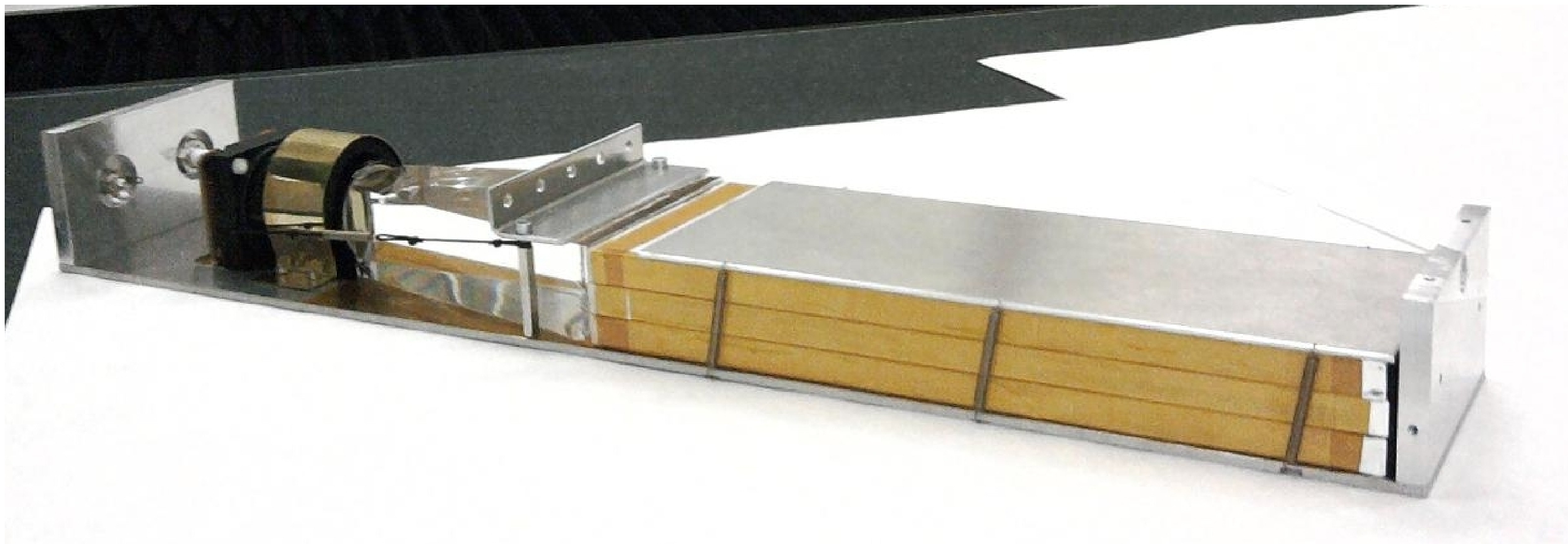}}~~
\boxed{\includegraphics[height=2.6cm]{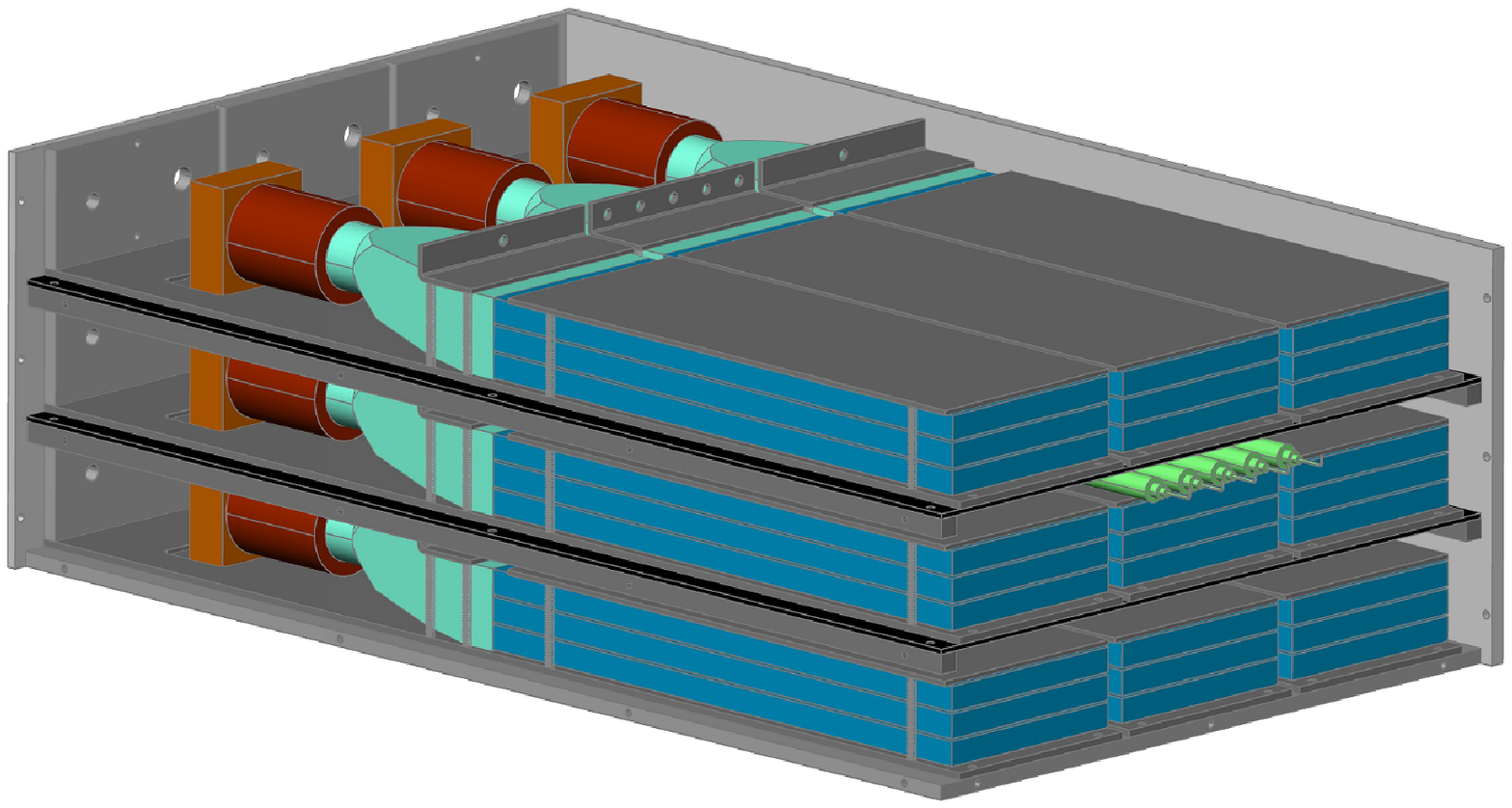}}
\end{center}
\caption{Left: a complete base module of the NEUCAL prototype, made of a stack of three scintillator layers connected to a photomultiplier through a Plexiglas light guide. Right: assembly of the prototype which has been tested at the CERN SPS in August 2009, made of a three by three stack of base modules provided of five $^3$He tubes.}
\label{fig_01_02}
\end{figure}

During 2009 a prototype NEUCAL detector has been produced to be tested on particle beams. The detector has a modular structure allowing the integration of three planes each made of three base modules. The single module is produced by assembling three 8.5\,cm\,$\times$\,25\,cm\,$\times$\,1\,cm fast scintillator layers, with a light guide in Plexiglas and a Hamamatsu fine mesh photomultiplier, model H5946. All scintillators, EJ-230 by Eljen Technology, have been cut and polished directly by the manufacturer. Light guides have been produced by INFN workshop in Florence and accurately polished in a laboratory of CNR by means of suspension of alumina powders in water and diamond powders in silicone oil. Each base modules has the possibility to be instrumented with up to five CANBERRA $^3$He proportional counter tubes, model 12\,NH25/1. The first module is shown in Fig. \ref{fig_01_02} (left) together with a drawing of the final configuration for the beam test (right). Five $^3$He tubes are installed between the first and second scintillator planes, in a symmetric position with respect to the detector lateral walls.

\section{Beam test}\label{sec:test}
\begin{figure}[top]
\begin{center}
\includegraphics[height=4.3cm]{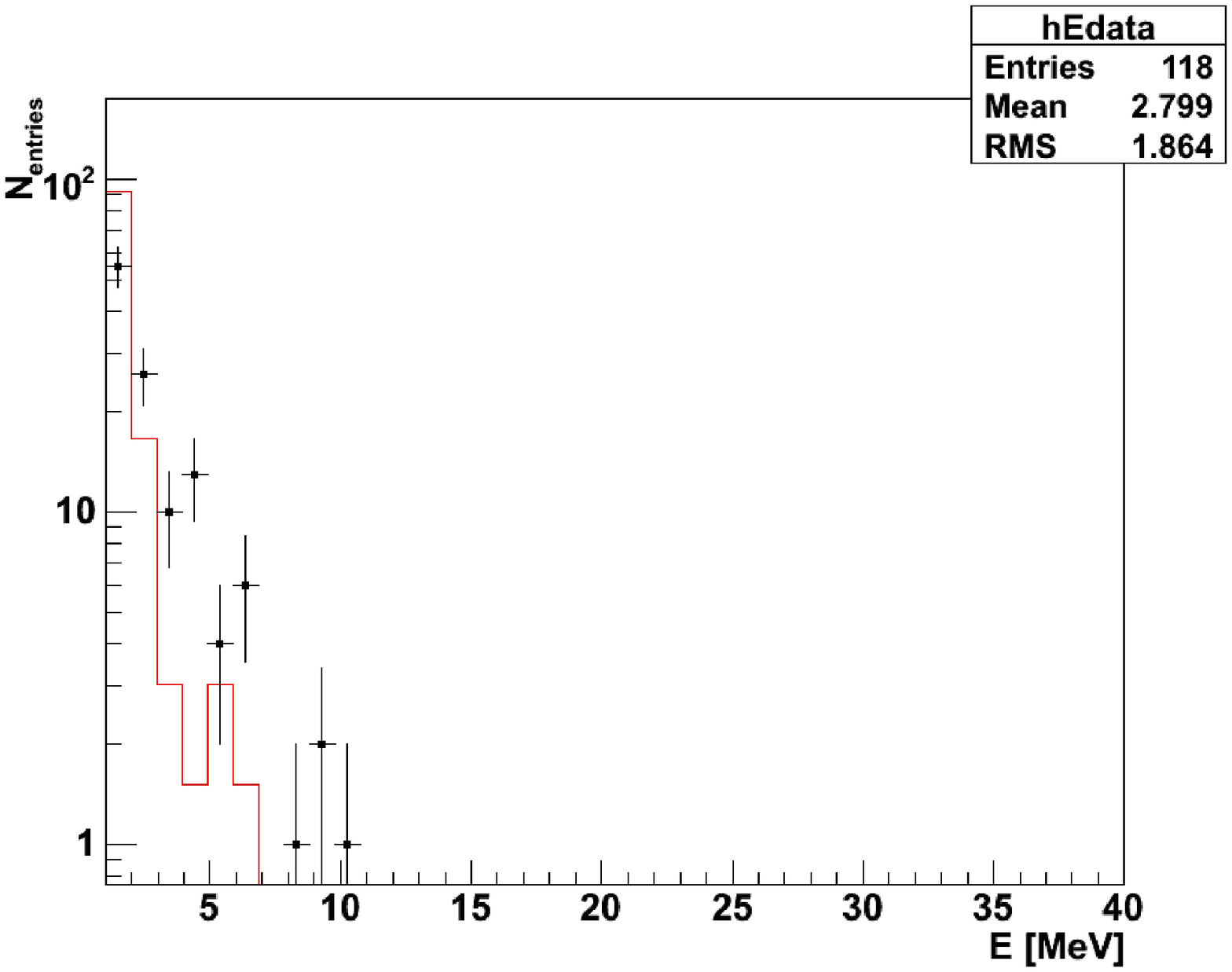}~~
\includegraphics[height=4.3cm]{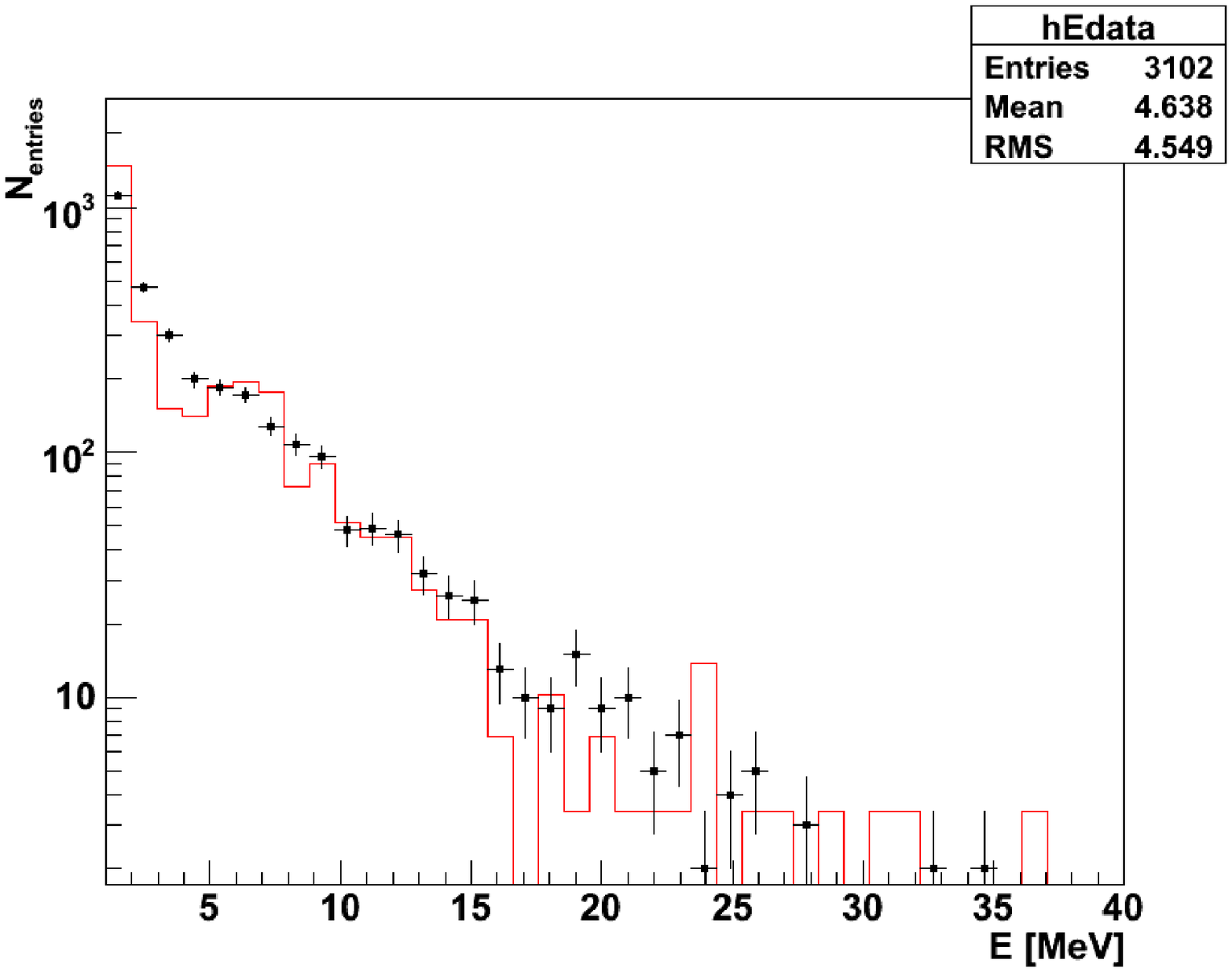}\\
\includegraphics[height=4.3cm]{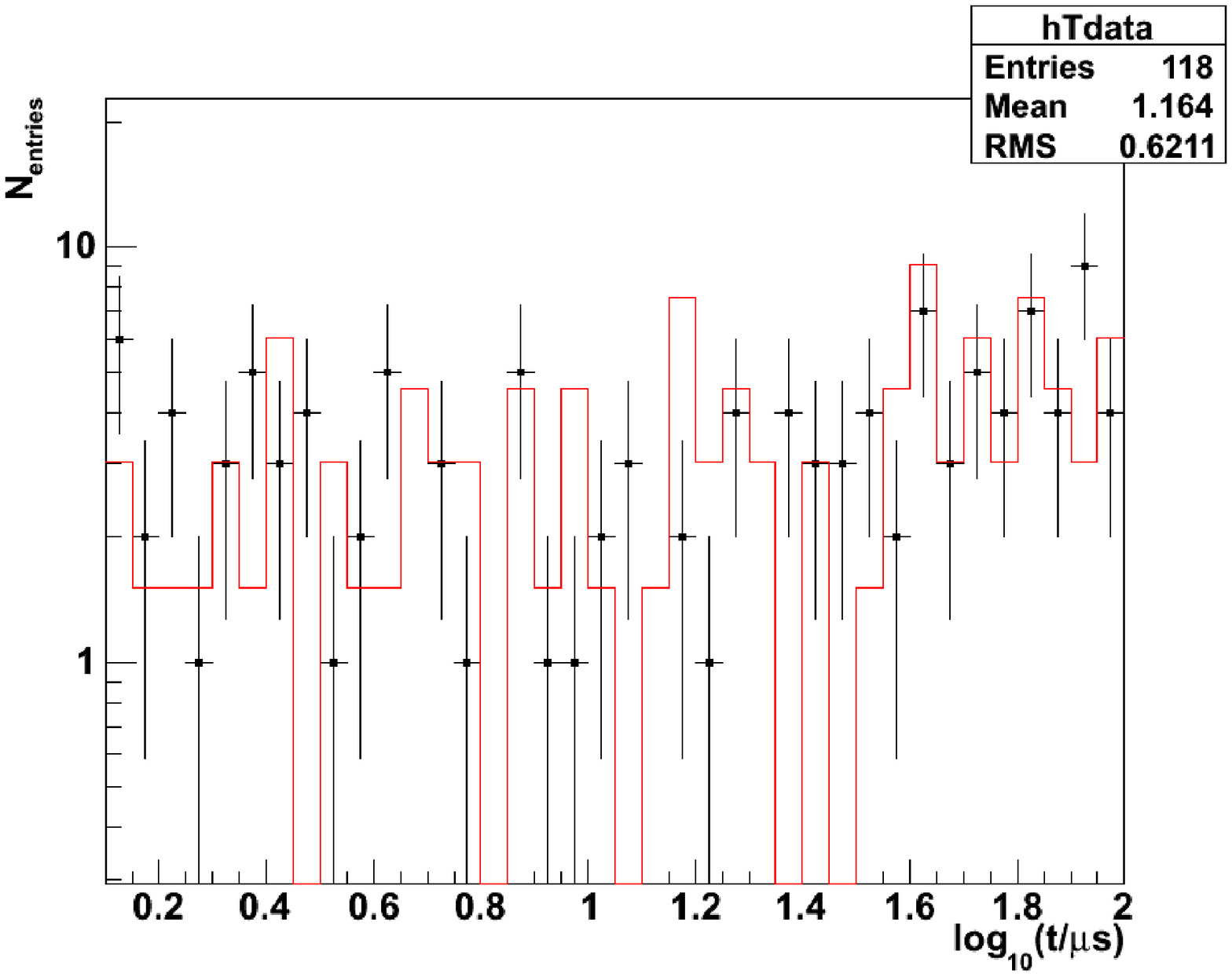}~~
\includegraphics[height=4.3cm]{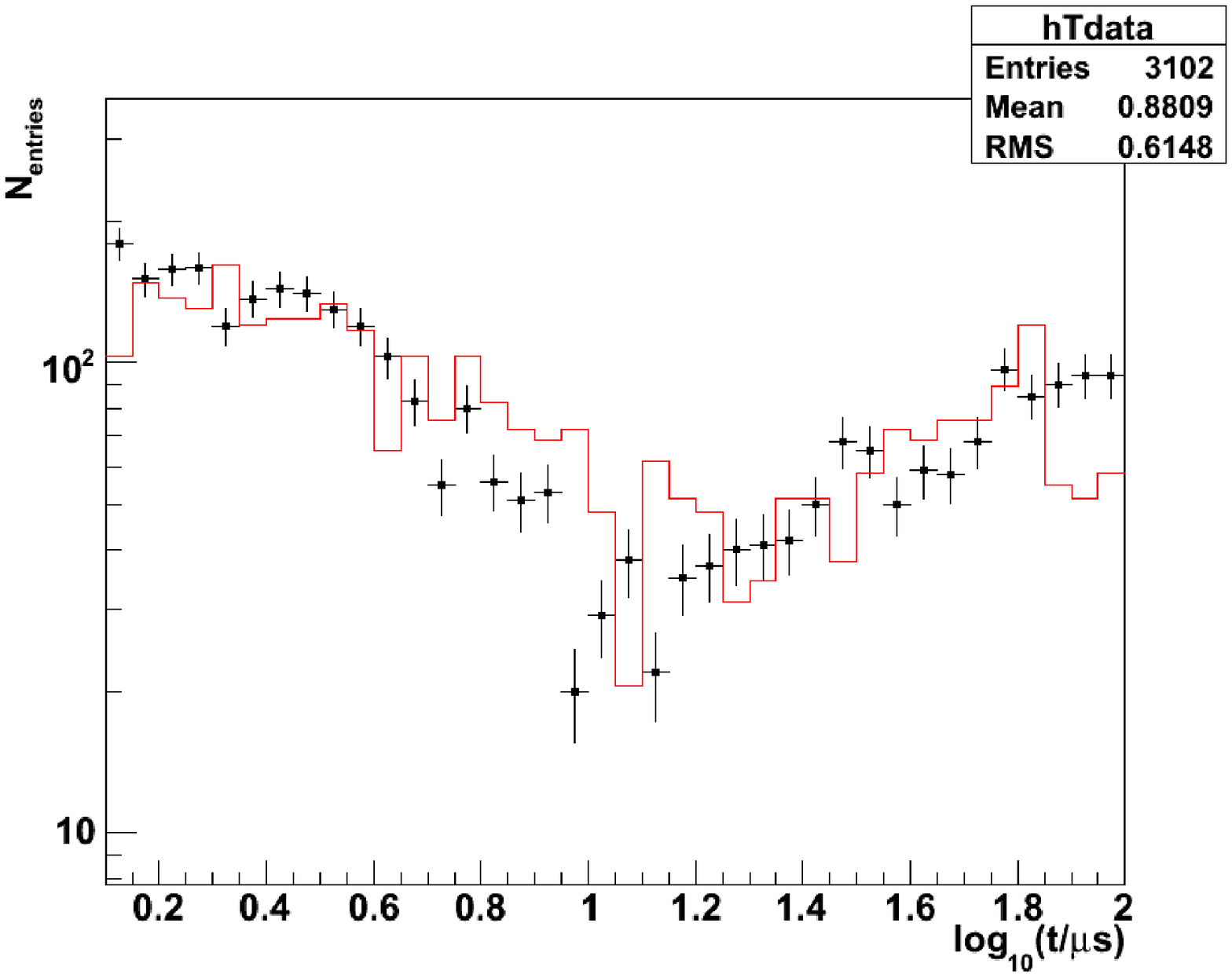}
\end{center}
\caption{Top: energy (top) and time (bottom) distribution of signals isolated in time for the central top NEUCAL scintillator module, in case of electron (left) and pion (right) showers.}
\label{fig_03_04_05_06}
\end{figure}

A beam test has been performed at the CERN SPS accelerator to study the response of the NEUCAL prototype to single minimum ionizing particles (MIPs) and to showers produced by electrons and hadrons. Muon, electron and pion beams have been used with energies between 100\,GeV and 350\,GeV. The aim of this test was mainly the identification and measurement of signals due to the scattering of neutrons inside the scintillator material following the development of e.m. and hadronic showers in an upstream calorimeter. FLUKA and GEANT4 simulations shows that the moderation (and possible subsequent capture) of neutrons with energy in the MeV region can take up to several hundreds microseconds. For this reason 1\,ms long sampling time was foreseen and achieved by means of two fast digitizer CAEN VME boards, model V1731 and V1720. For this test NEUCAL has been integrated together with a 16\,X$_0$ tungsten calorimeter, located in front of NEUCAL. Additional blocks of absorber material, made of lead tiles and lead tungstate (PbWO$_4$) crystals, could be added upstream in such a way to vary the calorimeter total depth. A very preliminary analysis has been performed for only one of the nine scintillator modules by using electron and pion samples taken in different configurations: 16\,X$_0$ and 29\,X$_0$ calorimeter depth respectively. Muon data have been used as a MIP sample for a preliminary energy calibration of the scintillator. Neutron signals have been identified as those signals coming after the complete shower development (during which we have intense signals on all scintillator modules) and isolated in time from all the other signals identified on scintillator modules and proportional counter tubes. Signals coming within few tens nanoseconds with respect to each other can be due to showers initiated by other particles of the beam or to single MIP particles crossing the detector. In figure \ref{fig_03_04_05_06} some preliminary comparison between data and a GEANT4 simulated events are shown. Arrival time of neutron signals has been restricted between 1\,$\mu$s and 100\,$\mu$s from the shower prompt signals. The four pictures show the signal energy (top) and time (bottom) distributions for electron (left) and pion (right) samples. A nice agreement can be seen which confirm that neutron signals are detected as expected, and that the neutron produced in hadronic shower are much more numerous than the neutrons produced in e.m. ones. Further complete analysis results will be published in a separate work.

\section{Conclusions}\label{sec:conclusions}
The NEUCAL neutron detector is under development for neutron identification in electron and hadron induced showers in e.m. calorimeters. It makes use of the novel concept of 'active moderation' and it is based on the detection of signals produced by elastic or inelastic scattering in a sensible moderator material during the moderation phase. A prototype detector has been tested at the CERN SPS with muon, electron and pion beams. The first very preliminary analysis shows that neutron moderation is feasible by means of plastic scintillators and detected signals due to neutron interactions are compatible with expectations in the time interval between 1\,$\mu$s and 100\,$\mu$s from the beginning of shower development. Further analysis is in progress.

\bibliographystyle{ws-procs9x6}
\bibliography{bonechi}

\end{document}